\documentclass [a4paper]{article}
\usepackage{amssymb}
\usepackage{amsfonts}
\usepackage{mathrsfs}
\usepackage{amsmath}
\usepackage{bbm}
\usepackage{indentfirst}
\usepackage{graphicx}

\begin{document}
    \renewcommand{\baselinestretch}{1.5}
    \centerline{\bf{\Large{An Improved Dilaton Background in Soft-Wall
    Model}}}\par\vspace{2em}
    \centerline{\large{Hongying Jin\footnote{Email: jhy@zimp.zju.edu.cn}, Gang Liu\footnote{Email: G\_Liu@zimp.zju.edu.cn}}}\par\vspace{1em}
    \centerline{\sl{Zhejiang Institute of Modern Physics, Zhejiang University, Hangzhou,
    China}}\par\vspace{2em}

    \centerline{\bf{\large{Abstruct}}}\par\vspace{1em}\normalsize
    We adopted the form of v.e.v. of the bulk scalar field proposed by P.Zhang\cite{pzhang} in this paper, and changed the
    dilaton background from $e^{-\kappa^2z^2}$ to $e^{-\kappa^2z^2+\alpha
    z}$ accordingly. The calculation shows that, as if we choose proper parameters,
    we can get a better result than both the present soft-wall models and the hard-wall model while
    keeping Regge behavior unchanged.\par\vspace{3em}

    \noindent {\Large\bf{1. Introduction}}\par
    So far there is no efficient way to perform analytical study on QCD
    in low-energy area, many efforts have been performed to approach this problem. Thanks to
    Maldacena's work\cite{maldacena}, it is realized that the information about four-dimensional
    strongly coupled gauge theories can be extracted from gravitational theories in higher
    dimensions. The AdS/QCD, the efforts apply a five-dimensional theory on an anti-de
    Sitter (AdS) gravity background to learn something about QCD, is one of such
    approaches, which draws wide attention.\par

    Although for QCD the exact form of the gravity dual is not yet
    known, the so-called bottom-up AdS/QCD approaches\cite{hardwall,softwall} still give us deep impression. In the so-called
    hard-wall model\cite{hardwall}, the authors introduce an infrared brane  to
    make the theory confining, and thus the bulk space becomes a slice of AdS
    space. This slice of AdS space certainly  matches QCD in the
    ultraviolet(UV)
    region due to QCD's asymptotic freedom. By choosing a "proper"
    cutoff, in ground hadron states, the hard-wall model fits the experiment
    well.\par
    For the high excited states, the hard-wall model could not describe the Regge behavior\cite{0602229}. In order to solve this
    problem, the so-called soft-wall
    model\cite{0602229} was proposed, which replaces the "hard" cutoff with a "soft" cutoff. By introducing an
    exponentially decaying background $e^{-\Phi}$, the expected Regge behavior can be recovered. It has been shown\cite{0602229} that, when the
    asymptotic behavior of $\Phi$ is $z^2$ as $z$ goes to infinity, the Regge behavior may be recovered. The simplest choice of the $\Phi$  is $\kappa^2z^2$,
    and the  corresponding calculation shows that\cite{softwall} it's much worse
    than the hard-wall model in the ground states. Some suggestions for improvement have been proposed\cite{imp1,ybyang,gkk,pzhang}.
    But generally these suggestions are not so satisfactory.\par
    Actually, the form of the $\Phi$ may be determined by the self-interaction of the scalar field $X$\cite{gkk}. In Ref.\cite{gkk}, where the
    self-interaction form is $X^4$, both the UV and IR behavior of $\Phi$ is $z^2$ and so the original soft-wall model is appreciated. However in Ref.\cite{pzhang},
    where the self-interaction form
    is $X^3$, the IR behavior of $\Phi$ is still $z^2$ while the UV behavior now becomes $z$. So in this work, we change the dilaton background
    $e^{-\kappa^2z^2}$ to $e^{-\kappa^2z^2+\alpha
    z}$, while adopting the form of v.e.v of the scalar form introduced in Ref.\cite{pzhang}. By choosing proper parameters, we can fit the physical results
    very close to the experimental ones. The average error between the theory and the experiment can be reduced to 5.77\%.
    Section 2 will explain our setup in detail, and we will show our results in section 3. In the last
    section we will take a discussion about our results.\par\vspace{2em}
    \noindent {\Large\bf{2. Setup}}\par
    In this section, we'll explain our modifications to the original soft-wall model in detail. The bulk space here is anti de Sitter space
    and the geometry takes the form of
    \begin{equation}
        ds^2=\frac{1}{z^2}(-z^2+\eta_{\mu\nu}dx^\mu dx^\nu)\text{,   }0\leq z \leq\infty
    \end{equation}
    And the bulk action in the theory is
    \begin{equation}
        S=\int d^5x\sqrt{g}e^{-\Phi}\left\{-\frac{1}{4g_5^2}\left(\|F_L\|^2+\|F_R\|^2\right)+\|DX\|^2-m_X^2\|X\|^2-\lambda\|X\|^3\right\}
    \end{equation}
    The $\|O\|^2$ signature means $\text{Tr}(O^\dag O)$ here. The bulk scalar field $X$, which is dual to $\bar{q}q$ and thus is responsible for the
    chiral symmetry-breaking, obtain a 5D mass $m_X^2=-3$ via
    the relation\cite{witten,polyakov} $(\Delta-p)(\Delta+p-4)=m^2$. Every 5D field corresponds to a 4D $p$-form operator and $\Delta$ here is the
    dimension of the operator. We add a cubic term as Ref.\cite{pzhang} suggested in the action. The covariant derivative of $X$ is
    $D^MX=\partial^MX-iA_L^MX+iA_R^MX$, where $A_{L,R}$ are chiral gauge fields and $A_{L,R}^M=A_{L,R}^{Ma}t^a$ with $t^a=\sigma^a/2$ being the generators
    of the gauged isospin symmetry. The gauge fields part remains unchanged and we'll focus on the scalar part.\par
    As in Ref.\cite{softwall}, the scalar field $X$ can be divided into two parts, the v.e.v $X_0(z)=v(z)\mathbbm{1}$ and the pion field $\pi^a$: $X=X_0(z)e^{2i\pi^a(x,z)t^a}$.
    Working in the axial-like gauge, $A_{L,Rz}=0$, we can get the e.o.m. of $v$:\par
    \begin{equation}
        \partial_z\left(\frac{e^{-\Phi}}{z^3}\partial_zv\right)-\frac{e^{-\Phi}}{z^5}\left(\frac 3 2\lambda v^2-3v\right)=0
    \end{equation}
    \noindent When $\lambda=0$, which reduces to the case of Ref.\cite{softwall} regardless of the dilaton form $\Phi$, we find $v=m_qz+\sigma z^3$
    in the UV region. However, in the IR region, this solution loses. Ref.\cite{pzhang} proposed one uniform form of $v(z)$:\par
    \begin{equation}
        v(z)=\frac{Az+Bz^3}{\sqrt{1+C^2z^2}}
    \end{equation}
    \noindent The parameters are determined by following relations
    \begin{equation}
        m_q=\frac{2A}{\zeta}\texttt{,    }\sigma=2\zeta\left(B-\frac12AC^2\right)\texttt{,    }\gamma=\frac{2B}{C}\texttt{,    }\zeta=\frac{\sqrt{3}}{2\pi}
    \end{equation}
    One can see that, when $z\rightarrow0$, $v$ behave as $(m_q\zeta z+\sigma z^3/\zeta)/2$ and when $z\rightarrow\infty$, $v\sim\gamma z^2/2$. According to
    the e.o.m., we can find the relation between the dilaton $\Phi$ and the v.e.v. $v$\cite{0752}:
    \begin{equation}
        \Phi'(z)=\frac{z^3}{v'}\left[\left(\frac{v'}{z^3}\right)'-\frac{1}{z^5}\left(\frac32\lambda v^2-3v\right)\right]
    \end{equation}
    We find that, in IR region, $\Phi\sim z^2$ which is just the requirement of the Regge behavior, and in the UV region, $\Phi\sim z$. So, we simply
    modify the dilaton background to $\Phi=\kappa^2z^2-\alpha z$.\par
    Adding an $\alpha z$ term is not just a mathematical game. Let's study the e.o.m. of the vector-like gauge field $V=A_L+A_R$:
    \begin{equation}
        \partial_z\left(\frac{e^{-\Phi}}{z}\partial_zV_{\mu n}^a\right)+\frac{m_n^2e^{-\Phi}}{z}V_{\mu n}^a=0
    \end{equation}
    We can denote $\psi_n=\left(\frac{e^{-\Phi}}{z}\right)^{1/2}V_n$. We have dropped the subscript $\mu$ out for convenience. We can thus rewrite
    the e.o.m. into the Schr\"{o}dinger form:
    \begin{equation}\label{kge}
        -\psi_n''+V(z)\psi_n=m_n^2\psi_n
    \end{equation}
    Here,
    \begin{equation}\label{potential}
        V(z)=\kappa^4z^2-\kappa^2\alpha z-\frac{\alpha}{2z}+\frac{3}{4z^2}-\frac{\alpha^2}{4}
    \end{equation}
    Actually, Eq.(\ref{kge}) is more like a Klein-Gordon equation rather than a Schr\"{o}dinger equation. Thus, $V(z)$ in Eq.(\ref{kge}) is somewhat like the square form
    of the potential in the corresponding Schr\"{o}dinger equation. In quark model, the potential of the Schr\"{o}dinger equation usually contains
    a linear potential, a Coulomb potential and vacuum energy. If such a potential writes like
    \begin{equation}
        V_S(z)=az-\frac{b}{z}+c
    \end{equation}
    We find that, its square form is
    \begin{equation}
        \tilde{V}(z)=a^2z^2+(2ac)z-\frac{2bc}{z}+\frac{b^2}{z^2}+c^2 - 2ab
    \end{equation}
    And such a potential is very similar to Eq.(\ref{potential}).
    \par\vspace{2em}
    \noindent {\Large\bf{3. Results}}\par
    Now we present our numerical results for QCD observables. There are four parameters in our model: $\kappa$,
    $\alpha$, $\gamma$ and the mass of quark $m_q$. The decay constant of pion $f_\pi$ and the mass of pion $m_\pi$ is fixed by experiment
    which are 92.3MeV and 139.6MeV separately.
    $\sigma$ can be determined by $m_\pi$, $f_\pi$ and $m_q$ through Gell-Mann-Oakes-Renner relation
    $m_\pi^2f_\pi^2=2m_q\sigma$ where $m_q$ is chosen in the region given by the Particle Data Book. The values of the other three parameters
    are chosen in order to fit the experimental results of QCD observables,
    and these values are:\par
    \begin{table}[htbp]
        \centering
        \begin{tabular}{ccccc}
            \hline\hline
            Parameter & $\kappa$ & $\alpha$ & $\gamma$ & $m_q$\\
            \hline
            Value(MeV) & 609.6 & 777.3 & 49.8 & 3.29\\
            \hline
        \end{tabular}
        \caption{Parameters Chosen in the Model.}
    \end{table}
    \noindent The results of QCD observables are exhibited in Tab.\ref{res}.
    \begin{table}[htbp]
        \centering
        \begin{tabular}{cccc}
            \hline\hline
            Observable & Experimental Value & Theoretical Value & Error(\%) \\
            \hline
            $m_\rho$ & $775.5\pm0.4$ & 775.5 & 0 \\

            $m_{a_1}$ & $1230\pm40$ & 1476 & 20 \\

            $f_{\rho}^{1/2}$ & $346.2\pm1.4$ & 344.7 & 0.43 \\

            $f_{a_1}^{1/2}$ & $433\pm13$ & 465 & 7.4 \\

            $g_{\rho\pi\pi}$ & $6.03\pm0.07$ & 6.09 & 1 \\
            \hline
        \end{tabular}
        \caption{Predictions for QCD observables, all experimental and theoretical values are in MeV, except $g_{\rho\pi\pi}$.}
        \label{res}
    \end{table}
    \noindent We find that, all results in ground state fit experiment very well except $m_{a_1}$, and the average error is about 5.77\%,
    which is much better compared with existing results. We also plot the electromagnetic pion form factor $F_\pi(Q^2)$ in the spacelike region
    below, and compare our results to the hard-wall model. One can see that our result is better, and fits the experiment quite well. Also, we
    can calculate the value of the mean pion charge radius squared through\cite{brodsky}\par
    \begin{equation}
        \left<r_\pi^2\right>=-6\left.\frac{dF_\pi(Q^2)}{dQ^2}\right|_{Q^2=0}
    \end{equation}
    \noindent The result is $0.456 fm^2$, which is consistent to the data from PDG\cite{pdg2010} $0.45 fm^2$.\par
    \includegraphics[width=12cm]{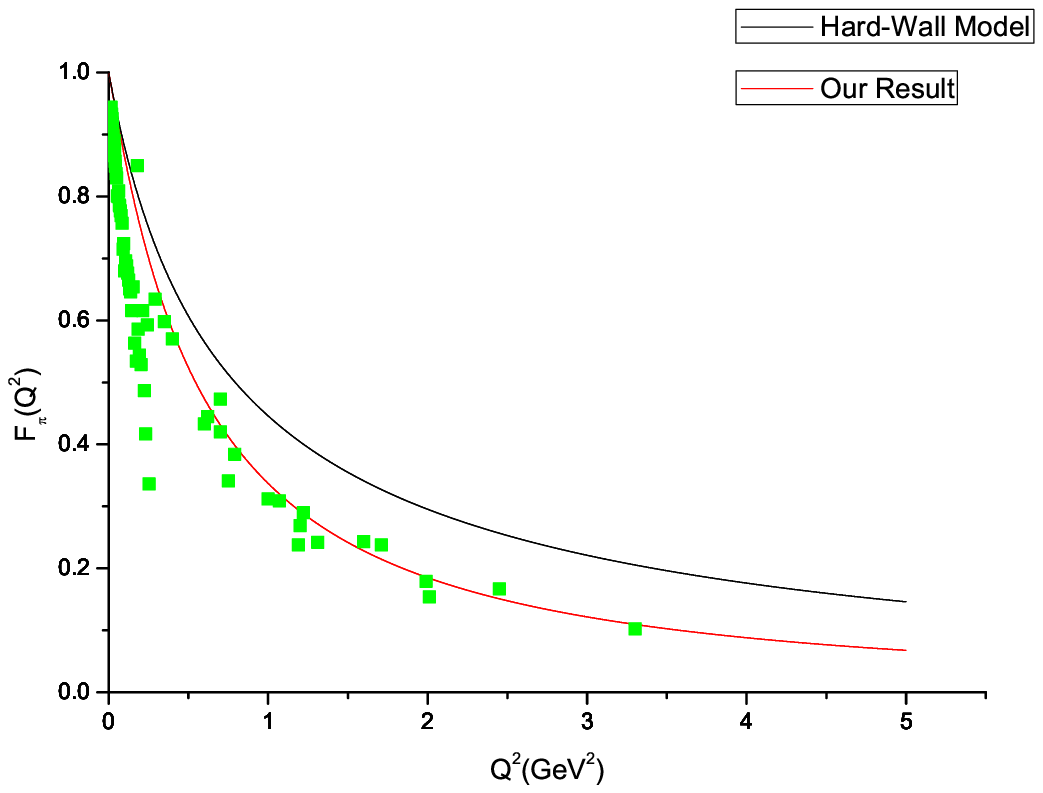}\par
    \noindent Figure 1: The electromagnetic pion form factor $F_\pi(Q^2)$ in the spacelike region. The blue plots are experimental results from
    Ref.\cite{exp}.\par\vspace{2em}
    \noindent {\Large\bf{4. Discussion and Conclusion}}\par
    The hard-wall model can get a satisfactory ground state result while violating the Regge behavior and the original soft-wall model can achieve the
    Regge behavior, but the ground state result is not so satisfactory. In this work, we reanalyzed the soft-wall model and proposed a new dilaton
    background. With these modifications, we get a better result than the hard-wall model while keeping Regge behavior unchanged.\par
    Honestly, the mass spectrum for both $\rho$ and $a_1$ is not so satisfactory. We list our result of $\rho$ here in comparison with the simplest
    soft-wall model in Ref.\cite{softwall}, the model in Ref.\cite{gkk} and the model in Ref.\cite{pzhang} in Tab.\ref{masspec}.\par
    \begin{table}[ht]
        \centering
        \begin{tabular}{ccccccc}
            \hline\hline
            $\rho$ & 0 & 1 & 2 & 3 & 4 & 5 \\
            \hline
            $m_{\texttt{exp}}$ & 775.5 & 1465 & 1570 & 1720 & 1900 & 2150 \\
            $m_1$ & 778 & 1100 & 1348 & 1556 & 1740 & 1906\\
            $m_2$ & 475 & 1129 & 1429 & 1674 & 1884 & 2072\\
            $m_3$ & 952.6 & 1349 & 1653 & 1909 & 2134 & 2338 \\
            $m_{\texttt{ours}}$ & 775.5 & 1376 & 1795 & 2149 & 2453 & 2715 \\
            \hline
        \end{tabular}
        \caption{Mass spectrum results. $m_{\texttt{exp}}$ are experimental results\cite{pdg2010}, $m_1$ are results from Ref.\cite{softwall},
        $m_2$ are from \cite{gkk}, $m_3$ are from Ref.\cite{pzhang},
        and $m_{\texttt{ours}}$ are our results. The average error of our results is 16.8\%.}
        \label{masspec}
    \end{table}
    The reason for the disappointing mass spectrum is probably that the model is too simple. Things are not difficult to
    find a group of parameters that fit the total mass spectrum, but then the ground state will not so satisfactory just like Ref.\cite{pzhang} whose
    ground state error is 22.8\%. Whether there is a model that can get satisfactory results for both ground state and the mass spectrum remains unknown.\par
     \noindent {\Large\bf{Acknowledgement}}\par
     This work is supported by NNSFC under Project No. 10775117.
    We would like to thank Prof. Jialun Ping for helpful discussions on numerical calculation.

\end{document}